# Drug-therapy networks and the predictions of novel drug targets

Zoltan Spiro, Istvan A. Kovacs and Peter Csermely

Address: Department of Medical Chemistry, Semmelweis University, P O Box 260., H-1444 Budapest 8, Hungary. Correspondence: Peter Csermely. Email: *csermely@puskin.sote.hu*

**Recently, a number of drug-therapy, disease, drug, and drug-target networks have been introduced. Here we suggest novel methods for network-based prediction of novel drug targets and for improvement of drug efficiency by analysing the effects of drugs on the robustness of cellular networks.**

Despite a significant and continuous increase in medical research spending, the number of new drugs approved by the US Food and Drug Administration (FDA), as well as the number of new drug targets each year remained rather constant in the past 20 to 25 years producing an annual increase of ~20 new drugs and ~5 new targets, respectively. Lengthy development procedures and the high risk of unexpected side-effects in advanced stage clinical trials decrease the innovativity of drug development. At this rate it will take more than 300 years to double the number of available drugs [1-4]. There are a number of ideas to overcome these burdens. Wide-range screens of existing drugs, seeking novel applications [2], combination therapy, i.e. the use of several drugs together or developing aptamer combinations [5-7], or the development of multi-target drugs [8-10] are all very promising areas of drug-design.

The organization of our rapidly growing knowledge on diseases, disease-related genes, drug targets and their structures as well as on drugs and pharmacologically relevant chemical structures gives us another exciting approach to discover novel drug-development areas. Besides data-mining [11] in 2007 and 2008 several networks have been constructed to help drug discovery [4,12-15]. In the network concept the complex system is perceived as a set of interacting elements, which are bound together by links. Links usually have a weight, which characterizes their strength (affinity, or propensity). Links may also be directed links, when one of the elements has a larger influence to the other than vice versa [16,17].

In a recent publication appeared in *BMC Pharmacology*, Nacher and Schwartz [14] compiled a drug-therapy network, where all US approved drugs and associated human therapies were connected to each other. They decomposed the original, bipartite network of therapies and drugs (where bipartite means that there are no connections *within* both therapies and drugs) into two networks. In the drug-network two drugs were connected, if there was at least one common therapy that was involved in both. Vice versa, two therapies were connected, if a drug implicated in both therapies existed. Their analysis provided the first view of the relationships between therapies as defined by drug-therapy interactions highlighting a few key drugs, which connect distinct therapy classes in a few steps.

Both the drug and therapy networks proved to be small-worlds [16,17], i.e. distant therapies were separated by less than three chemicals as an average. Highly connected therapies, therapy-hubs should play a relevant role in the drug-therapy network, because the network behaved close to a tree-like network, where the relative importance of hubs is high. The majority of drugs (79%) were grouped in clusters connected to a specific therapy. However, there was a minority of drugs (21%), which formed bridges spanning different therapeutic classes. These drugs may acquire a particular significance. In order to characterize the main set of drugs and therapies that governs the network, Nacher and Schwartz [14] computed a several network centrality measures. They identified a sub-network of the above-mentioned bridging drugs with high betweenness centrality in the drug-therapy network including *Scopolamine*, *Morphine*, *Tretinoin* and *Magnesium Sulfate*. As an example, *Tolbutamide* and *Magnesium Sulfate* defined a key shortest path of two steps between distant therapies like "Insulines and analogues" (A10) and "Dermatological preparations" (D11). Apparently unrelated disorders were thus separated by a much lower number of chemicals than could be expected. A majority of drugs act on one target, but a small number of drugs act on a large number of targets. Nacher and Schwartz [12] proposed a special role for those drugs, which have a high betweenness centrality *and* act on multiple targets. They especially highlighted *Hydroxocobalamin*, *Vitamin B3*, *Vitamin B12*, *Atropine*, *Ophenadrine* and *Procaine*.

The network approach not only gives us a systematic way to organize our vast databases, but also provides a visual image, which mobilizes the integrative power of our right brain hemisphere to understand the daunting complexity of these systems. However, many networks such as that of the 1,360 individual chemical substances studied by Nacher and Schwartz [14] are often too big for easy visualization. The Anatomic Therapeutic Chemical (ATC) classification system used by these authors [14] gave them an excellent possibility to construct a hierarchical representation of drug-therapy information, where we may zoom-in from the top layer of 15 anatomical main therapeutic groups, through the 66 therapeutic subgroups ($2^{nd}$ layer), the 123 pharmacological subgroups ($3^{rd}$ layer), the 448 chemical subgroups ($4^{th}$ layer) until reaching the $5^{th}$ layer of 1360 individual chemical substances. Where the data-set is not as straightforwardly hierarchical as the ATC-classification [14], network hierarchy can be explored by various other techniques [18-20].

To show the rich context, where the results of the Nacher and Schwartz paper [14] can be rationalized and used, we will extend the above information and (1) overview the powerful network approach to construct various drug-target and related networks proposing several novel network construction methods; (2) suggest a number of novel approaches to predict new drug targets by network analysis, and (3) propose the regular assessment of the robustness of human cellular networks by simulating the network perturbations of drug-candidates to avoid unwanted resistance and side effects. These novel, system-based analytical tools may significantly extend the powerful approach of the drug-therapy network set by Nacher and Schwartz [14], and have a great promise to increase the number of novel drug targets and improve the approval rate of new drugs.

**Drug-target and related networks**
When thinking on the possible network representations of diseases, drugs and drug targets (for a comprehensive list of the available databases see Table 1), first the elements of the network have to be defined. As the next step a general rule should be found determining both the elements linked in the particular network and the nature (weight, directedness) of the links connecting them. Besides the already mentioned drug-therapy network [14], recently a number of other network-building rule-sets have been published [4,12,13,15], which give additional exciting and novel information on the vast datasets of diseases, drugs and drug targets. A summary of these representations is shown on Figure 1.

The pharmacological space linking drugs and other protein ligands to their targets has been explored by four recent publications [4,12,13,15]. In these approaches the network can be constructed by either linking two drug target proteins, if both bind one or more compound, or by linking two compounds (drugs), if there is at least one protein as their common target. It is rather sobering for highly creative medicinal chemists that the average molecular weight becomes smaller and smaller as we go from preclinical drug-candidates to Phase I, II, III and approved drugs. Moreover, other physicochemical properties, like hydrophobicity and H-bonding potentials further reduce the available chemical space for the mostly desired oral drugs [13]. The analysis of the drug/target-network reveals further elements of the low-risk behaviour of the pharmaceutical industry. The network is particularly enriched in highly targeted proteins and elements with many neighbours (called hubs) are preferentially connected to each other forming a so-called 'rich-club'. This is a result of the tendency to target an already validated target protein with alternative or follow-up compounds. Experimental drugs reveal a much wider 'exploration radius', if assessed by both the diversity of target proteins and the localization of the targets elsewhere than the preferential plasma membrane. However, so far these efforts have not led to a significant expansion of novel targets meaning novel protein classes and novel cellular compartments [12].

An additional approach to decipher meaningful information for drug development efforts is to link those human diseases, which have at least one common genes involved in the development of the disease. This human disease network was also converted to the complement-network of disease genes, where two genes are connected, if they are associated with the same disorder. Among human diseases several types of cancer, such as colon and breast cancer are hubs that are genetically connected to more than 30 distinct disorders. Disease genes that contribute to a common disorder often (i) have protein products forming larger complexes; (ii) are often co-expressed and (iii) have similar major functions. Interestingly, those inheritable disease genes, which are not essential, occupy a peripheral position in the cellular network. This is in stark contrast to essential genes, which are more central. On the contrary, disease genes associated with somatic mutations such as somatic cancer genes have a central position in cellular networks [21]. When comparing drug-target networks with the related diseases, an ongoing shift of drug development can be observed towards 'novel diseases' with associated genes that were not prior drug targets [12].

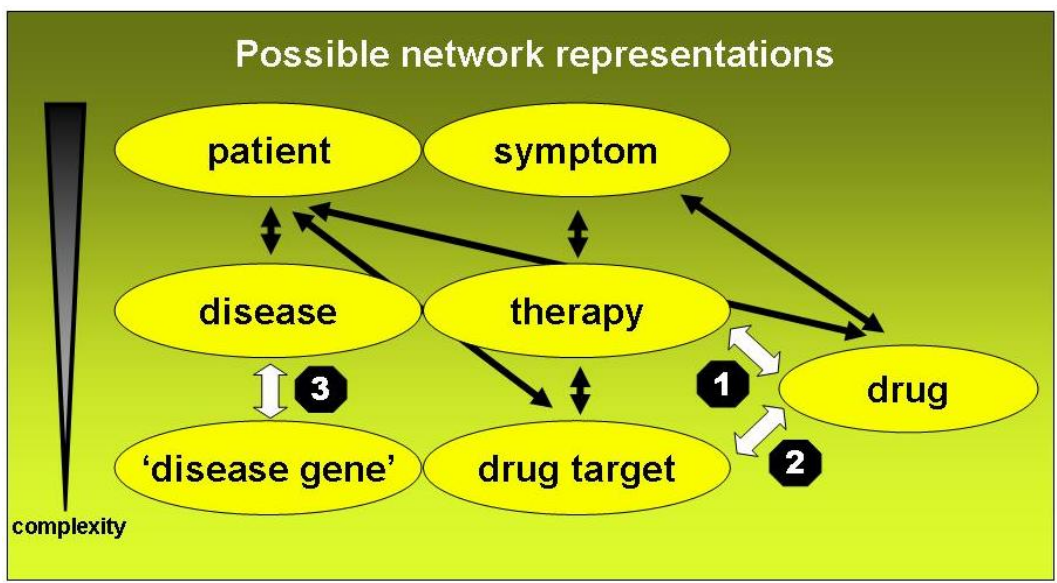

**Figure 1** Overview and possible extensions of therapy, disease, drug and drug-target networks. The figure summarizes the data-types (datasets), which were already used for network construction (connected with white arrows) or may be used to construct similar networks in the future (connected with black arrows). Datasets are positioned from top to bottom in the approximate order of their decreasing complexity. An arrow represents a network construction method, where two elements of the dataset at the end of the arrow are linked, if both of them are related to the same element(s) of the dataset at the start of the arrow. Names of network structures marked with numbers are the following: (1) the drug-therapy network of Nacher and Schwartz published recently in *BMC Pharmacology* [14]; (2) drug-target networks [4,12,13,15]; (3) disease-gene network [21]. Datasets positioned in the same row are overlapping with each other. Please note that the list of the datasets on this figure can be grossly extended. We need a lot more data analysis to show, or disprove the usefulness of the network approach at each of the data combinations in understanding the complexity of the available disease and drug information.

In the analysis and visual representation of drug-therapy and drug/target networks the weight of the links (e.g. the number of drugs binding to both targets in the drug target network) is seldom assessed. Additionally, these networks have not been thoroughly analyzed by defining their groups, or modules [16-20]. Both additions will certainly provide more detailed information on these exciting datasets. Important messages could be drawn from the additional networks described on Figure 1. Not only drugs, but their respective drug targets can also be linked to the various therapies. As an additional, rich source of data, patient records can be analyzed by the diagnosed diseases as well as prescribed drugs. Patient medication records can be transcribed to a patient-drug target network, which may reveal novel aspects of the phenotype-variability of diseases. Yet an additional set of data lie in the symptoms of patients, which can serve as a basis to construct symptom-disease, symptom-therapy or symptom-drug networks (Figure 1).

Drugs may also form a structural network, where two drugs are linked, if they contain the same, signature-like chemical segment or feature. Drugs can also be assembled to form a side-effect network, or toxicity-network, which may give an overall view of these two key maladies of drug development. As more and more data will be available in the future, patient symptoms can be extended by appropriately selected patient transcriptome, proteome, metabolome, oral microbiome and gut microbiome data. Here we have to warn that we do not think that the proposed 'inflation' of drug and drug-related networks will solve the current problems of drug-design. Just oppositely, the more networks we add, the less clarity and focus we may enjoy. However, drug- and disease-related network representations will certainly have their own evolution, and we will have to see, which of them will give us the most straightforward, non-obvious visual and analytical information. It is our great pleasure to invite our colleagues to explore this rich field, and provide the Darwinian triad of (i) mutations, (ii) proliferation and (iii) natural selection of drug and drug-related networks.

**Prediction of novel drug targets using network analysis**
Existing segments of drug-target networks may have hidden information on additional drug targets, which are yet not included to the network. Extension of existing networks is a recent, exciting field of network studies. It may be worth to review these approaches in hope to suggest appropriate measures to predict future targets from the currently available targets and other data-sets. We may predict both links and elements in networks. Link-prediction is a much more straightforward procedure, thus we start with a summary of these methods [22,23]. Missing links can be predicted from
a.  the similarity of their putative two endpoints (the more we know from the inherent properties of the network elements, the better predictions we can make) [24,25];
b.  the similarity of their neighbourhood in the network (this neighbourhood may be restricted to their common neighbours, or may include all their neighbours, even the second neighbours, the whole group they belong to, or the whole network) [23,26,27];
c.  the comparison of the actual network to an appropriately selected model network [23,28], which has been addressed excellently by the recent Nature paper of Newman and colleagues [19];
d.  the analysis of sequential snapshots of network topology (also called as network dynamics, or network evolution) [29].

In the above list, the appropriate measure of "similarity" is a central and rather difficult problem of mathematics, physics, ecology and many other scientific disciplines. Finding a similarity index becomes notoriously difficult, when the similar structures are complex, such as the topology of network segments in point "b". Application of the Jaccard's coefficient or other commonly used similarity indices [23] can provide an appropriate solution to this problem. However, the use of machine-learning methods [24,25,27], more complex similarity measures [20,28] or both can be even more beneficial. An alternative solution to the same problem is to extend *bona fide* drug target networks (where the network elements are drug targets) by novel elements. These novel elements may occupy structural holes [30] of the original network, or may be proposed by methods, like chance-discovery [31].

The identification of novel drug target candidates can be accomplished by finding missing links in all networks, where drug targets serve as links, e.g. in drug [4,12,13,15] therapy, or patient networks arranged by joint drug targets. The above link and node discovery methods can extend each other in *all* bipartite networks, where any of the disjoint group of elements (often called as 'top' or 'bottom' elements) can be converted to links joining the other set of elements. This may give us novel methods to predict, discover, test and extend gene regulatory, metabolic, opinion (recommendation), collaboration (co-authorship), sexual and any other affiliation-type networks. As a closing remark of this section, we have to note that a 'multi-neighboured' element of a bipartite network having more than one joint element in the other set of the bipartite network corresponds to a clique and not to a single link in the transformed regular network. This feature requires a 'collection-step' of adjacent links to a clique, when predicting such, multi-neighboured elements.

**Prediction of efficient drug-candidates overcoming the robustness of cellular networks**
Robustness is an intrinsic property of cellular networks that enables them to maintain their functions in spite of various perturbations [32]. (We have to note that this 'dynamical' definition of robustness differs from the 'structural' definition of robustness, where the same word often refers to the maintenance of network connectivity. 'Structural' robustness is more appropriately called as network persistence [17].) Networks of different topology vary by orders of magnitude in their robustness to mutations and noise. Enhanced robustness is a property of only a very small number of all possible network topologies [33]. Cellular networks both in health and in disease belong to this extreme minority and show a differently shaped robust behaviour (Figure 2).

Many times when a drug fails or produces side effects, cellular robustness provides most of the explanation. A drug can be ineffective, when the robustness of cellular networks of disease-affected cells or parasites compensates for its changes. On the contrary, drug side effects can be the result of hitting an unexpected point of robustness-fragility of the affected networks [34]. Robustness analysis is already used to reveal primary drug targets [35], and the first methods have also been established to give a quantitative measure of changes in robustness during drug action [36].

Cellular robustness can be caused by a number of mechanisms.
a.  Strong links (meaning intensive, high probability, high affinity interactions [17]) often form negative or positive feedbacks helping the cell to return to the original state (attractor) or jump to another, respectively. This systems control enables the system to move between two stable states.

b.  The contribution of weak links (meaning non-intensive, low probability, low affinity interactions [17]) is more diffuse. Weak links provide (i) alternative, redundant, degenerate pathways; (ii) flexible connections disjoining network modules to block perturbations and re-assembling them in a slightly altered fashion, and (iii) additional, yet unknown mechanisms buffering the effects of the original perturbation further, and de-coupling physical perturbations from functional-level activities [17,32,34]. We must note that weak links grossly outnumber strong links in cellular networks, which may often make the effects of their fail-safe mechanisms larger than those of negative or positive feedbacks.
c.  Finally, robustness of cellular networks is also helped by an increased average robustness of their elements (e.g. proteins), which, most of the time are networks by themselves [17].

Cellular robustness is linked to evolvability, thus the potential of cells to evolve. There is an intricate balance in the cell to preserve the general stability by robustness and to allow an evolutionary change by an increased evolvability.

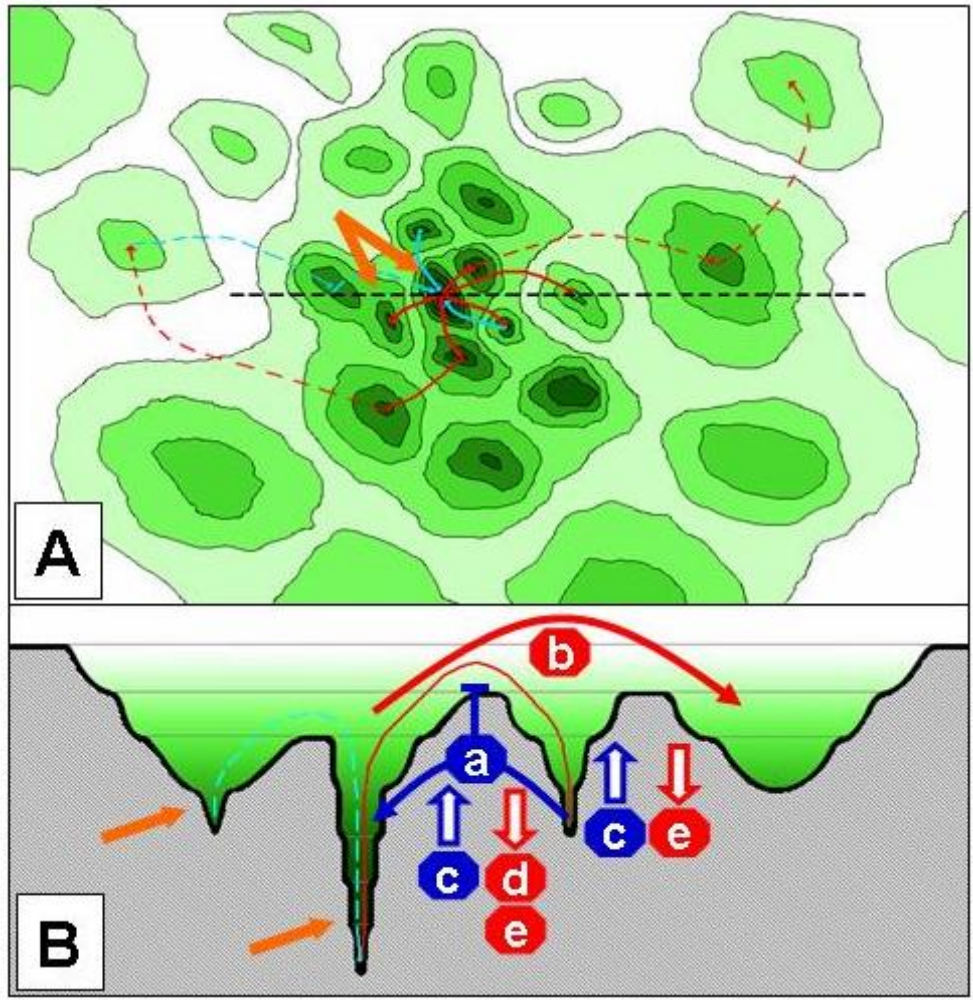

**Figure 2** An illustrative sketch of the stability-landscape of 'healthy' and 'diseased' phenotypes and possible ways to overcome the robustness of the system by drug action. Panel A describes a 2-dimensional contour plot of the stability landscape of 'healthy' (the central and the adjacent two minima marked with the orange arrows) and 'diseased' phenotypes (all additional local minima). Darker green colours refer to states with larger stability. Red arrows are examples of 'bad' changes from 'healthy' to one of the 'diseased' states, while blue arrows represent 'good' changes either from a diseased to a healthy state or between two healthy states. Dashed arrows refer to less probable changes. Panel B illustrates the various modes of drug action on cellular robustness. The valleys and hills are a vertical representation of the stability-landscape shown on Panel A along the horizontal dashed black line. Blue symbols represent drug interactions with disease-prone or diseases-affected cells, while red symbols refer to drug effects on disease-affected parasites. Letters denote the various cases described in the text in detail. (a) counteracting decoy or new regulatory feedback; (b) positive feedbacks pushing the diseased cell or parasite to another

trajectory; (c) blockade of the general destabilization of cells; (d) hit of the Achilles-heel; (e) 'error-catastrophe' [20,32,34].

The achievement of robustness always has a price. Robust cells have their fragile point, their 'Achilles-heel', and can not be optimized to all other aspects of cellular life, such as proliferation. This gives us chances to conquer or re-direct cellular robustness by the application of drugs (Figure 2). We may develop drugs, which

a. act like a counteracting decoy (a Trojan-horse) or establish a new regulatory feedback helping the disease-affected cell to return the original equilibrium;
b. hit one or more positive feedbacks pushing the diseased cell or parasite to another trajectory – in most cases this has to be accompanied by a general dampening of buffering weak links;
c. block the general destabilization of cells in its transition towards a diseased state or during aging by enhancing cellular buffering;
d. find the Achilles-heel of the cellular robustness of disease-affected cells or parasites;
e. decrease the effect of buffering weak links to the extent that the disease-affected cell or parasite suffers an 'error-catastrophe', where robustness collapses [17,32,34].

Here we have to note that the decreased robustness of cells in transition towards a diseased state and that of aging cells, as well as the purposeful decrease of robustness we would like to achieve by our drug action on disease-affected cells or parasites makes the cellular network noisier and less predictable. Thus the better job our 'c' to 'e'-type drugs make from the above list, the more difficulties we may encounter to find them by currently available, 'equilibrium' network analytical methods. The development of 'fuzzy', stochastic network analysis [20,28] as well as the comparison of network time-series (which also allow us to take into account the possibilities of circadian chronotherapy, and examine other cellular dynamics [34]) may help us to overcome this difficulty.

As a take-home message of this section we want to stress that our current knowledge on cellular networks and their analytical methods arrived to the point, which allows the test of the effects of drug candidates with known cellular targets or target-sets on the robustness of cellular networks. This "robustness-test" revealing both resistance-related failures and a number of side effects should be a mandatory element of standard drug-development protocols. The more we will know on tissue- and disease-specific changes of cellular networks and on the individual variations of these changes, the better we will predict the efficiency of drugs in *in silico* experiments.

In summary, we have shown that the recently published drug-therapy [14] and drug-target networks [4,12,13,15] as well as their potential extensions (Figure 1) provide a powerful and exciting tool for the organization of the expanding drug-development data as well as give us a global view on major trends and limitations. We also raised the idea that the recent advances in link and node discovery methods [22-31] give an excellent chance for the suggestion and verification of novel drug targets. Finally, we warned that drugs and their combinations should always be tested for their effects on the robustness of cellular networks. The advent of combinatorial therapies [5-7] and multi-target drugs [8-10] may significantly help us to break or re-direct the robust behaviour of the cell. However, for the knowledge-based design of appropriate drug combinations and multi-target drugs we need a number of novel approaches and techniques to explore the dynamical complexity of cellular networks after multiple perturbations. We gave a summary of these multi-target drug design methods elsewhere [37,38].

**Addendum**
Recently Campillos et al. [39] published a network of 502 drugs and their side effects using these data to predict novel drug targets based on side-effect similarity of two chemically dissimilar drugs. From the same data a side-effect network could also be constructed, where two side effects are linked, if a drug exists, which has both. This side effect network in combination with the link-prediction methods outlined above [19,22-29] opens the possibility to predict additional side effects of existing drugs and drug candidates. Network-based side effect prediction would greatly help the development of better clinical trial protocols, and would uncover additional possible dangers before the large-scale use of novel drugs. In another recent paper Lee et al. [40] described a disease network, where links represented mutated enzymes and connected diseases showed increased co-morbidity.

**Acknowledgements**
Authors would like to thank the members of the LINK-Group (www.linkgroup.hu) for their helpful comments. Work in the authors' laboratory was supported by the EU (FP6-518230) and the Hungarian National Science Foundation (OTKA K69105). The funding bodies had no role in the content and submission of this manuscript.

**Table 1.** Useful links to therapy, disease, drug and drug-target network data

| Name and description | Link | Reference |
| --- | --- | --- |
| **DrugBank:** a bioinformatics-cheminformatics resource combining detailed drug data with comprehensive drug target information with >4900 drug (~3500 experimental) and >1500 non-redundant protein entries | http://www.drugbank.ca/ | [41] |
| **Drug-Target Network:** network data of 890 drugs and 394 target human proteins | http://www.nature.com/nbt/journal/v25/n10/suppinfo/nbt1338_S1.html | [12] |
| **Drug-Therapy Network:** three layers of drug-therapy networks according to the ATC classification | http://www.biomedcentral.com/1471-2210/8/5/additional/ | [14] |
| **FDA Orange Book:** approved drug products with therapeutic equivalence evaluations | http://www.fda.gov/cder/ob/ | |
| **IDdb:** Thomson Investigational drugs database including information on 107000 patents, 25000 investigational drugs and 80000 chemical structures | http://scientific.thomson.com/products/iddb/ | |
| **OMIM:** a knowledgebase of human genes and genetic disorders | http://www.ncbi.nlm.nih.gov/sites/entrez?db=omim | [42] |
| **PDTD:** 3D drug target structure database with a target identification option | http://www.dddc.ac.cn/pdtd/ | [43] |
| **Predicted drug targets:** a set of 1383 predicted drug targets | http://www.biomedcentral.com/1471-2105/8/353/additional/ | [25] |
| **Protein ligand network:** a network of 4208 ligands and ~15000 binding sites | http://pbil.kaist.ac.kr/~parkkw/Lnet/ | [15] |
| **TDR Targets Database:** identification and ranking targets against neglected tropical diseases | http://tdrtargets.org/ | |
| **Therapeutic Target Database:** lists >1500 therapeutic targets, disease conditions and corresponding drugs | http://xin.cz3.nus.edu.sg/group/cjttd/ttd.asp | [44] |